\documentclass{aa}
\usepackage{natbib}
\usepackage{amssymb,amsmath}
\usepackage{longtable,enumerate}
\usepackage{rotating,afterpage}

\def\he{helium}

\def\fe{iron}

\def\chandra{\emph{Chandra}}
\def\xspec{\emph{XSPEC}}
\def\apj{\texttt{ApJ}}

\def\chandra{\emph{Chandra}}
\def\apj{\texttt{ApJ}}
\def\apec{\texttt{APEC}}

\begin{document}

\title{The Effect of Helium Sedimentation on Galaxy Cluster Masses and Scaling Relations}

\author{
G.~E.~Bulbul\inst{\ref{inst1}, \ref{inst2}}
\and N.~Hasler\inst{\ref{inst2}}
\and M.~Bonamente\inst{\ref{inst2},\ref{inst3}}
\and M.~Joy\inst{\ref{inst3}}
\and D.~Marrone\inst{\ref{inst4}}
\and A.~Miller\inst{\ref{inst5}}
\and T.~Mroczkowski\inst{\ref{inst6},\ref{inst7}}
}

\institute{Harvard-Smithsonian Center for Astrophysics, Cambridge, MA 02138\label{inst1}
\and Department of Physics, University of Alabama, Huntsville, AL 35899 \label{inst2}
\and NASA Marshall Space Flight Center, Huntsville, AL 35812 \label{inst3}
\and Department of Astronomy, University of Arizona, Tucson, AZ
85721 \label{inst4}
\and  Department of Physics, Columbia University, New York, NY 10027 \label{inst5}
\and Einstein Postdoctoral Fellow \label{inst6}
\and Department of Physics and Astronomy, University of Pennsylvania, Philadelphia, PA 19104 \label{inst7}
}

\abstract
{Recent theoretical studies predict that the inner regions of galaxy clusters may have an enhanced helium abundance due to sedimentation over the cluster lifetime. If sedimentation is not suppressed (e.g., by tangled magnetic fields), this may significantly affect the cluster mass estimates.}
{We use \chandra\ X-ray observations of eight relaxed galaxy clusters to
investigate the upper limits to the effect of helium
sedimentation on the measurement of cluster 
masses and the best-fit slopes of the $Y_{X}$-$M_{500}$ and $Y_{X}$-$M_{2500}$ 
scaling relations.}
{We calculated gas mass and total mass in two limiting cases: a uniform, unenhanced abundance distribution
and a radial distribution from numerical simulations of helium sedimentation on
a timescale of 11 Gyrs.}
{The assumed helium sedimentation model, on average, produces a negligible increase in the gas 
mass inferred within large radii ($r< r_{500}$)  (1.3 $\pm$ 1.2 \%) and a 10.2 $\pm$ 5.5  \% mean
decrease in the total mass inferred within $r< r_{500}$.
Significantly stronger effects in the gas mass (10.5 $\pm$ 0.8 \%) and total mass (25.1 $\pm$ 1.1 \%) 
are seen at small radii owing to a larger variance in \he\ abundance in the inner region, $r \leq0.1\, r_{500}$.}
{We find that the slope of the $Y_X-M_{500}$ scaling relation is not significantly affected
by helium sedimentation.}


\maketitle

\section{Introduction}
\label{intro}

Clusters of galaxies are permeated by  an optically thin ionized 
plasma at a temperature of $\sim 10^{8}$ K. This hot 
intra-cluster medium (ICM) is composed primarily
of hydrogen and helium ions, with a small fraction
($<$1 \%) of heavier elements, such as iron.
A commonly adopted assumption is that
\he\ and other heavy elements are uniformly distributed
throughout the ICM, but theoretical studies initiated 
by \citet{fabian1977}, \citet{rephaeli1978} and \citet{abramopoulos1981}
suggest that diffusion processes will result in 
enrichment of \fe\ and other heavy elements 
in the cluster core. Furthermore,
\citet{gilfanov1984}, \citet{qin2000}, \citet{chuzhoy2003} and \citet{chuzhoy2004}
have shown that the diffusion speed is greater
for lighter elements, and conclude that the diffusion of 
\he\ ions into the core of galaxy clusters -- helium sedimentation --
can occur within a Hubble time. 
However the problem of \he\ sedimentation in galaxy clusters is still under debate.
Recent theoretical studies show that the presence of thermal diffusion \citep{Shtykovskiy2010} 
and strong magnetic fields \citep{peng2009} in the intra-cluster medium may suppress sedimentation
of \he\ ions into the core of clusters.

The diagnostics of abundance profiles in the intra-cluster plasma
are based primarily on line emission from heavy elements. 
Since helium is fully ionized in clusters and produces no lines in the X-ray energy band, 
its radial distribution and abundance cannot be constrained by X-ray spectroscopic data alone. 
X-ray observations combined with observations of the Sunyaev-Zel'dovich effect (SZE)
may be used to infer the \he\ abundance in the intra-cluster medium \citep{markevitch2007}; 
however, the limited resolution and sensitivity of current SZE interferometers 
have so far not enabled the measurement of the distribution of \he, even in the cluster cores
where the sedimentation effect is the strongest. New capabilities for high angular resolution, high sensitivity SZE observations are currently being developed (e.g. at the Combined Array for Research in Millimeter-wave Astronomy, CARMA) which will allow observational constraints to be placed on helium sedimentation in cluster cores.

The sedimentation of \he\ ions within clusters of galaxies
may affect the physical quantities derived from X-ray 
observations such as cluster gas mass, total mass 
\citep{qin2000, ettori2006}, and the cosmological parameters derived from
these quantities \citep{markevitch2007,peng2009}.
Using high signal to noise \chandra\ observations of eight relaxed galaxy clusters,  
we investigate a theoretical upper limit to the effect of \he\ sedimentation on the 
measurement of cluster properties; in particular, we study how this impacts mass 
determinations and the $Y_{X}$-$M_\Delta$ scaling relations within overdensities 
$\Delta=2500$ and $\Delta=500$. 
To determine the upper limit of \he\ sedimentation effect we use a
limiting case of a \he\ sedimentation profile for 11-Gyr-old
galaxy clusters, based on the numerical simulations provided
by \cite{peng2009}. This work is the first application of a theoretically
predicted \he\ sedimentation profile to \chandra\ observations.

This paper is organized as follows: in \S \ref{section:modeling} 
we provide the details of our data reduction and modeling. 
The impact of \he\ sedimentation on X-ray derived gas and total mass
and on the $Y_{X}$-$M_\Delta$ scaling relation is presented and discussed 
in \S \ref{section:heliumSedimentation}.
In \S \ref{section:conclusion} we provide conclusions. 
In all calculations we assume the cosmological parameters
$h=0.73$, $\Omega_M=0.27$ and $\Omega_{\Lambda}=0.73$.

\begin{table}[h!t]{
\begin{centering}
\scriptsize
\caption{Cluster Sample}
\vspace{1 mm}
\begin{tabular}{lcccccc}
\hline
\hline
\\
Cluster 		& z				& $N_{H}$ $^{a}$ 		& Obs. ID& Exp. Time\\ 
			&				& (cm$^{-2}$)		&	& (ksec)\\\hline
\\
MACS~J0744.9+3927	& 0.69				& 5.6$\times 10^{19}$	& 6111	& 49.5 \\
			&				&			& 3585	& 19.7 \\
			&				&			& 3197	& 20.2 \\
\\
Zwicky~3146		& 0.29				& 2.5$\times 10^{20}$	& 909	& 46.0 \\
			&				&			& 9371	& 38.2 \\
\\
MACS~J1311.0-0311	& 0.49				& 1.8$\times 10^{20}$	& 3258	& 14.9 \\
			&				&			& 6110	& 63.2 \\
			& 				& 			& 9381	& 29.7 \\
\\
Abell~1835		& 0.25				& 2.0$\times 10^{20}$	& 6880	& 110.0	\\
\\
MACS~J1423.8+2404	& 0.54				& 2.2$\times 10^{20}$	& 4195	& 115.6 \\
\\
MACS~J1621.6+3810	& 0.46				& 1.1$\times 10^{20}$	& 3254	& 9.9 \\
			&				&			& 6109	& 37.5 \\
			&				&			& 6172	& 29.8 \\ 
			&				&			& 9379	& 29.2 \\
			&				& 			& 10785 & 29.8 \\
\\

Abell~2204 		& 0.15				& 5.7$\times 10^{20}$	& 7940	& 72.9	\\
\\
MACS~J1720.3+3536	& 0.39				& 3.5$\times 10^{20}$	& 3280	& 20.8	\\
			&				&			& 6107	& 33.9\\
\\	
\hline  
\hline
\label{table:clusterSample}
\end{tabular}\\
$(a)$ Leiden/Argentine/Bonn (LAB) Survey, see \citet{kalberla2005}\\
\vspace{5 mm}
\end{centering}
}
\end{table}

%%%%%%%%%%%%%%%%%%%%%%%%%%%%%%%%%%%%%%%%%%%%%%%%%%%%%%%%%%%%%%%%%%%%%%%%%%%%%

\section{X-ray Observations and Modeling}
\label{section:modeling}

Our goal is to obtain density and temperature profiles for the gas
in clusters based on \chandra\ X-ray imaging and spectroscopy,
using a cluster model which accounts for
a variable \he\ abundance as a function of radius.
We describe below the reduction of the X-ray data,
the application of the cluster model, and the measurement of
cluster gas mass and total mass.

\subsection{X-ray Data Reduction}
\label{section:dataReduction}

We chose the eight galaxy clusters from the \citet{allen2008} sample
with the deepest \chandra\ ACIS-I observations (unfiltered integration time 
$\geq 60$ ksec), all of which are available as \chandra\ archival data 
(see Table \ref{table:clusterSample}).  The sample spans a range of redshift 
(0.15 $<$ z $<$ 0.69).
As part of the data reduction procedure,
we apply afterglow, bad pixel and charge transfer inefficiency corrections to the 
Level 1 event files using CIAO 4.1 and CALDB 4.1.1. 
We use the \citet{markevitch2003} method for light curve filtering 
to eliminate flares in the background due to solar activity. 
The exposure times after filtering 
are given in Table~\ref{table:clusterSample}.
We follow the method described in \citet{bulbul2010}
for background subtraction.

We use the 0.7-7.0 keV energy band extraction of spectra and images in order 
to minimize the effect of calibration uncertainties at the lowest energies,
and the effect of the high detector background at high energy.
Spectra are extracted in concentric annuli surrounding the X-ray emission centroid.
In this process X-ray point sources were excluded. An optically 
thin plasma emission model (\apec) is used, with temperature, abundance 
and normalization as free parameters in the spectroscopic fit (in \xspec). 
The redshifts $z$ and Galactic neutral hydrogen column densities $N_{H}$ of the 
clusters are shown in Table~\ref{table:clusterSample}.
We adopt a 1\% systematic uncertainty in each bin of the surface 
brightness and a 10\% systematic uncertainty in temperature profile 
\citep{bulbul2010}.       
The products of this reduction process
are the X-ray surface brightness and temperature profiles
shown in Figure \ref{fig:fits}.

\begin{figure}[!h]
\centering
\includegraphics[angle=0,width=8cm]{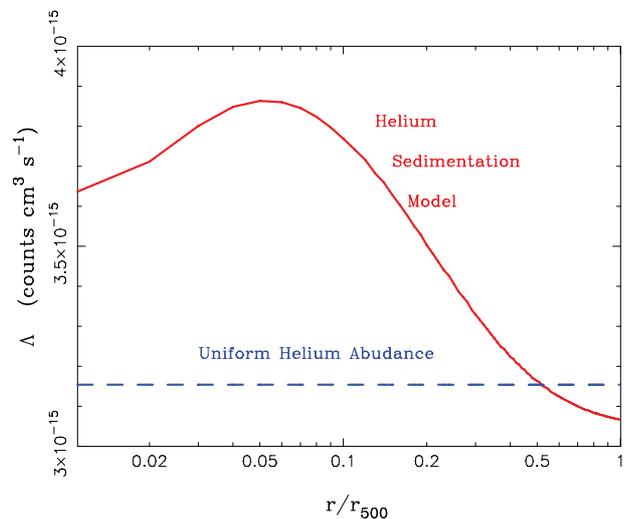}
\caption{The band-averaged X-ray cooling function, $\Lambda_{ee}(r)$, 
for two different \he\ distributions in the ICM: a uniform 
\he\ abundance (blue dashed line) and a \he\ sedimentation distribution provided 
by \citet{peng2009} (red line).
The overdensity radius $r_{500}$
is the radius within which the mean cluster 
density is a factor of 500 times greater than the critical density
of the universe at the cluster redshift.}
\label{fig:bandAveragedEmissivity}
\vspace{5 mm}
\end{figure}

\subsection{Application of Cluster Models to X-ray Data}
\label{sec:emissivity}

We start with the projected surface brightness
\citep{birkinshaw1991,hughes1998}:
\begin{equation}
S_{x}= \frac{1}{4\pi(1+z)^{3}} \int n_{e}^{2}\ \Lambda_{ee}\ d\ell ,
\label{eqn:surfaceBrightness}
\end{equation}
where $S_{x}$ is in detector units (counts s$^{-1}$ cm$^{-2}$ arcmin$^{-2}$), 
$\Lambda_{ee}$ is the band averaged  X-ray cooling function in 
counts cm$^{3}$ s$^{-1}$,
$n_{e}$ is the number density of electrons, 
\textit{z} is the redshift, and the integral is along the line of sight, $\ell$.

We calculate the X-ray cooling function in two limiting cases: (i) a uniform
abundance \citep{anders1989} and (ii) the radial abundance profile of \cite{peng2009}
obtained for a sedimentation time scale of 11~Gyr with no magnetic field. 
We use the astrophysical plasma emission database (APED) which includes thermal 
bremsstrahlung, radiative recombination, 
line emission, and two-photon emission \citep{smith2001}.
In addition, we apply corrections for galactic 
absorption \citep{morrison1983}, relativistic effects
\citep{itoh2000}, and electron-electron bremsstrahlung \citep{itoh2001},  
to obtain the X-ray emissivity ($\epsilon_{\nu}$) as a function of frequency $\nu$.

The band-averaged X-ray cooling function ($\Lambda_{ee}(T,r)$) is obtained 
by integrating the $\epsilon_{\nu}$
over the observed \chandra\ energy band (0.7-7.0 keV):

\begin{equation}
\Lambda_{ee}(T,r) = \frac{\int_{\nu}\epsilon_{\nu} d\nu}{n_{e}^{2}}.
\label{eqn:apecEmissivityIntegrated}
\end{equation}

\noindent The resulting band-averaged X-ray cooling function for a uniform
abundance \citep{anders1989} and a \he\ sedimentation
profile \citep{peng2009} is shown in Figure \ref{fig:bandAveragedEmissivity}.

\begin{figure*}
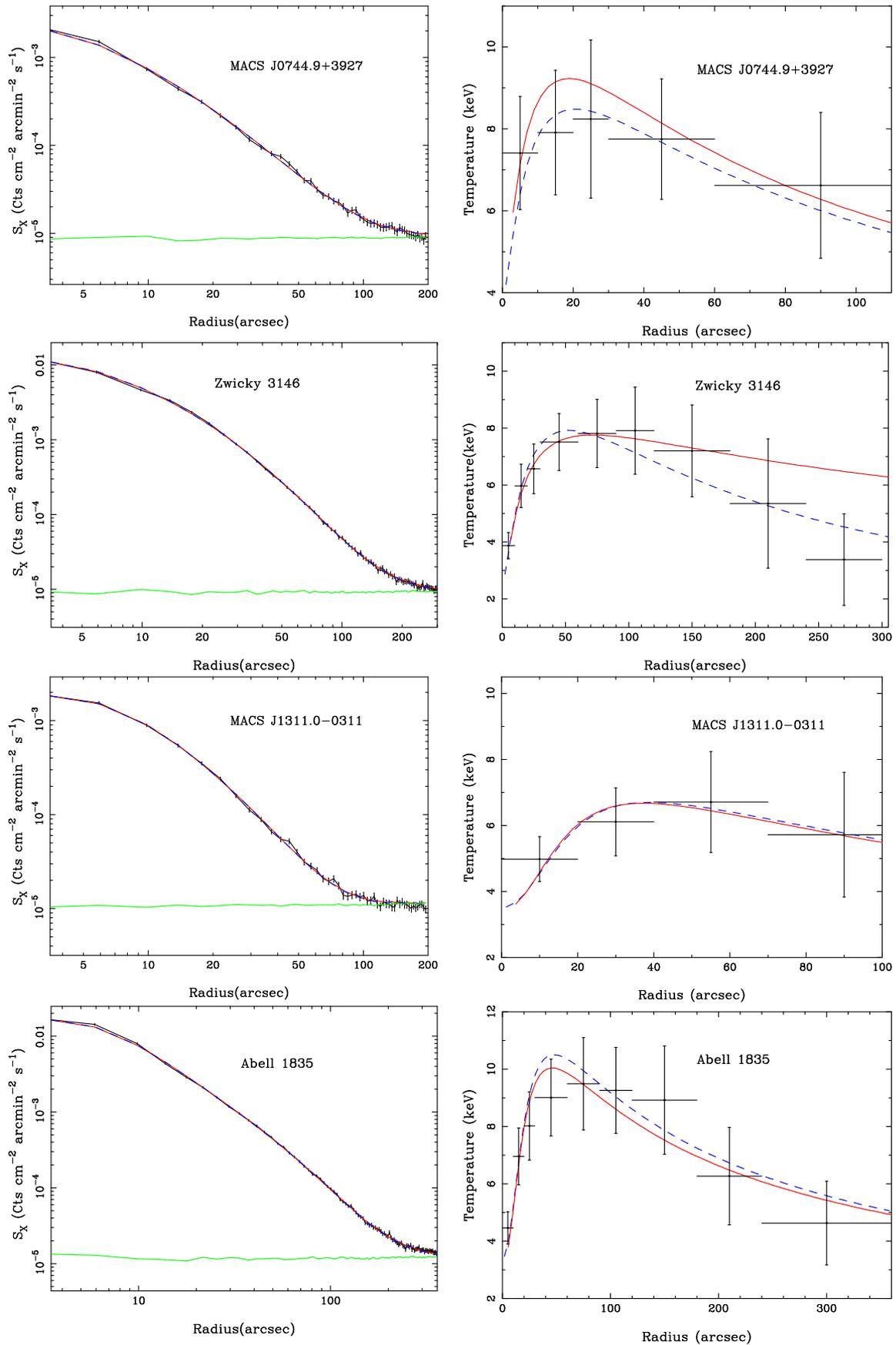

\begin{center}
\includegraphics[angle=-90,width=0.4\textwidth]{f1a_new.eps}
\hspace{0.01\textwidth}
\includegraphics[angle=-90,width=0.4\textwidth]{f1b_new.eps}
\newline
\includegraphics[angle=-90,width=0.4\textwidth]{f1c_new.eps}
\hspace{0.01\textwidth}
\includegraphics[angle=-90,width=0.4\textwidth]{f1d_new.eps}
\newline
\includegraphics[angle=-90,width=0.4\textwidth]{f1e_new.eps}
\hspace{0.01\textwidth}
\includegraphics[angle=-90,width=0.4\textwidth]{f1f_new.eps}
\newline
\includegraphics[angle=-90,width=0.4\textwidth]{f1g_new.eps}
\hspace{0.01\textwidth}
\includegraphics[angle=-90,width=0.4\textwidth]{f1h_new.eps}
\newline
\caption{X-ray surface brightness (left column)  and temperature profiles (right column) for the sample 
of eight relaxed galaxy clusters.  The black points are derived from the X-ray image and spectroscopic data. 
The blue lines in both profiles show the best fit model to the data obtained 
using a uniform \citet{anders1989} \he\ abundance, and the red lines show 
the best fit model obtained using
the \citet{peng2009} \he\ sedimentation distribution; the green lines in surface brightness 
profiles indicate the background levels determined from the blank sky observations. The overall $\chi^{2}$ of the fits
obtained from the uniform \he\ abundance and the \he\ sedimentation model
(Tables~\ref{table:bestFitParams_A&G} and \ref{table:bestFitParams_P&N}) show that 
both distributions give equally good fits to \chandra\ data.}
\label{fig:fits}
\end{center}
\end{figure*}

\begin{figure*}
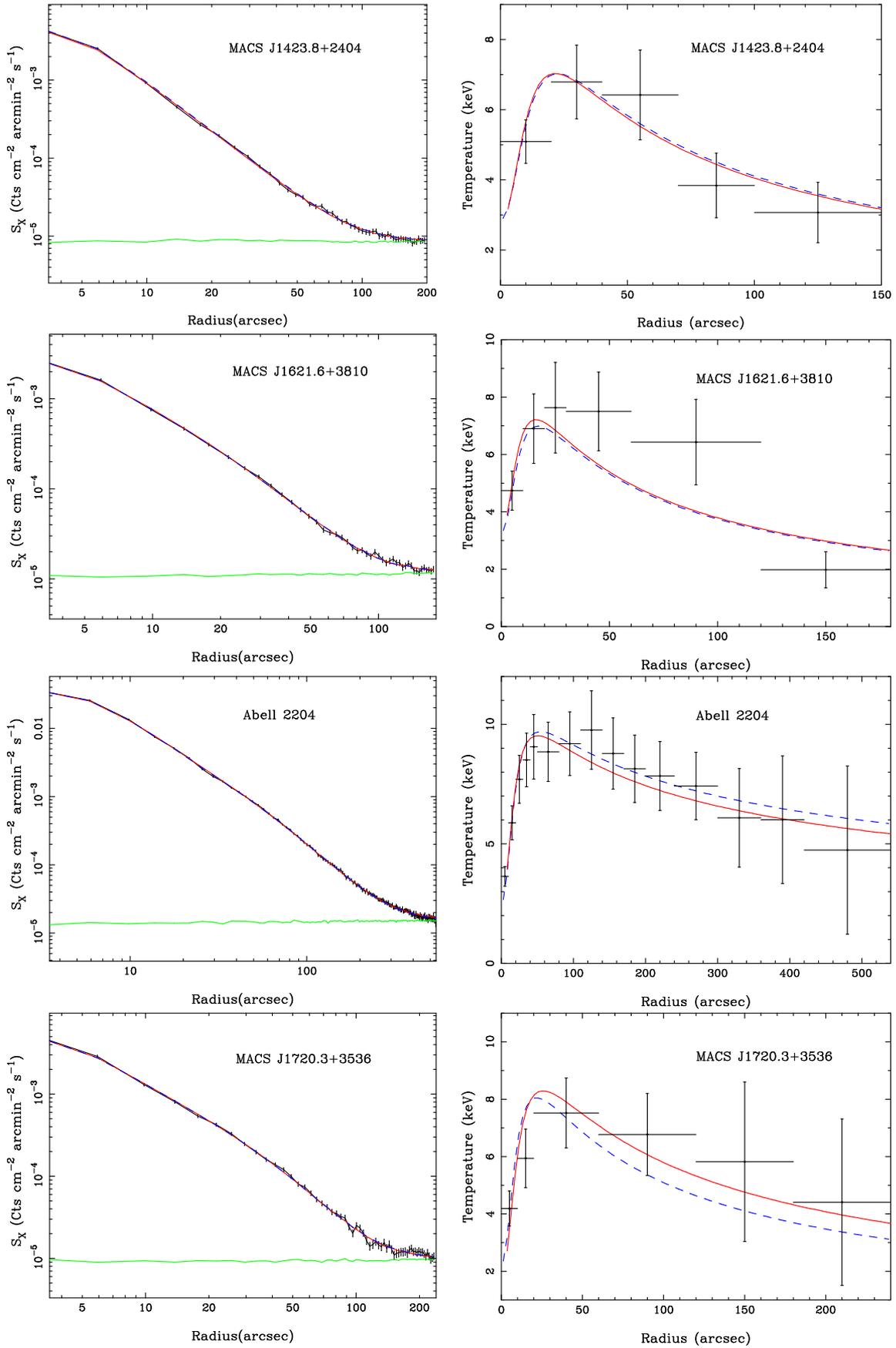

\begin{center}
\includegraphics[angle=-90,width=0.4\textwidth]{f1i_new.eps}
\hspace{0.01\textwidth}
\includegraphics[angle=-90,width=0.4\textwidth]{f1k_new.eps}
\newline
\includegraphics[angle=-90,width=0.4\textwidth]{f1l_new.eps}
\hspace{0.01\textwidth}
\includegraphics[angle=-90,width=0.4\textwidth]{f1m_new.eps}
\newline
\includegraphics[angle=-90,width=0.4\textwidth]{f1n_new.eps}
\hspace{0.01\textwidth}
\includegraphics[angle=-90,width=0.4\textwidth]{f1p_new.eps}
\newline
\includegraphics[angle=-90,width=0.4\textwidth]{f1r_new.eps}
\hspace{0.01\textwidth}
\includegraphics[angle=-90,width=0.4\textwidth]{f1s_new.eps}
\newline
\addtocounter{figure}{-1}
\caption{X-ray surface brightness and temperature profiles for our sample 
of galaxy clusters (cont'd).}
\end{center}
\end{figure*}

To obtain the electron number density $n_{e}(r)$ and temperature 
$T(r)$ profiles, we apply the analytic model for the
intra-cluster plasma developed by \citet{bulbul2010}
and compare the model predictions to the observed surface 
brightness and temperature data points found in \S \ref{section:dataReduction}.

The electron number density is
\begin{equation}
\begin{aligned}
n_{e}(r)=& n_{e0} \left(\frac{1}{(\beta-2)}\frac{(1+r/r_{s})
^{\beta-2}-1}{r/r_{s}(1+r/r_{s})^{\beta-2}} \right)^{n}\tau_{cool}^{-1}(r)\\
\end{aligned}
\label{eqn:coolcore_gas_density}
\end{equation}
where $n_{e0}$ is the central value of the number density of electrons, 
$r_{s}$ is the scaling radius, $\beta+1$ is the slope of the total
matter density profile, and \textit{n} is the polytropic index. $\tau_{cool}(r)$
is the phenomenological core-taper function used for cool-core clusters
\citep{vikhlinin2006},
\begin{equation}
\tau_{cool}(r)=\frac{\alpha+(r/r_{cool})^{\gamma}}{1+(r/r_{cool})^{\gamma}}.
\label{eqn:cooling_funct}
\end{equation}
The corresponding \citet{bulbul2010} temperature profile is 
\begin{equation}
\begin{aligned}
T(r)=&T_{0}\left(\frac{1}{(\beta-2)}\frac{(1+r/r_{s})^{\beta-2}-1}{r/r_{s}(1+r/r_{s})^{\beta-2}} \right)\tau_{cool}(r)\\
\end{aligned}
\label{eqn:coolCore_tempProf}
\end{equation}
where $T_{0}$ is the normalization parameter, and the other parameters
are in common with those in Equation \ref{eqn:coolcore_gas_density}. 

We vary the model parameters using a Monte Carlo Markov Chain
approach \citep{bonamente2004,bulbul2010};
the best fit model parameters for a uniform \he\ abundance and a
\he\ sedimentation scenario are given in Table \ref{table:bestFitParams_P&N} 
and are shown by the solid blue and red curves in Figure \ref{fig:fits}.
We assume that clusters with redshift z $>$ 0.3 have
$\beta=2$, since the polytropic
index (\textit{n}) and $\beta$ cannot be determined simultaneously
from the data available.
For all of the clusters we use a core taper  
parameter $\gamma=2.0$ except for Zwicky~3146,
for which we obtained a better fit with $\gamma=1.0$.
With these choices of fixed parameters, we obtain acceptable fits
to the data for both the assumed uniform and sedimented \he\ profiles
(see Tables \ref{table:bestFitParams_A&G} and \ref{table:bestFitParams_P&N}).

\subsection{Cluster Mass}
\label{sec:clusterMass}

\begin{figure}[hbp!]
\centering
\includegraphics[angle=-90,width=0.45\textwidth]{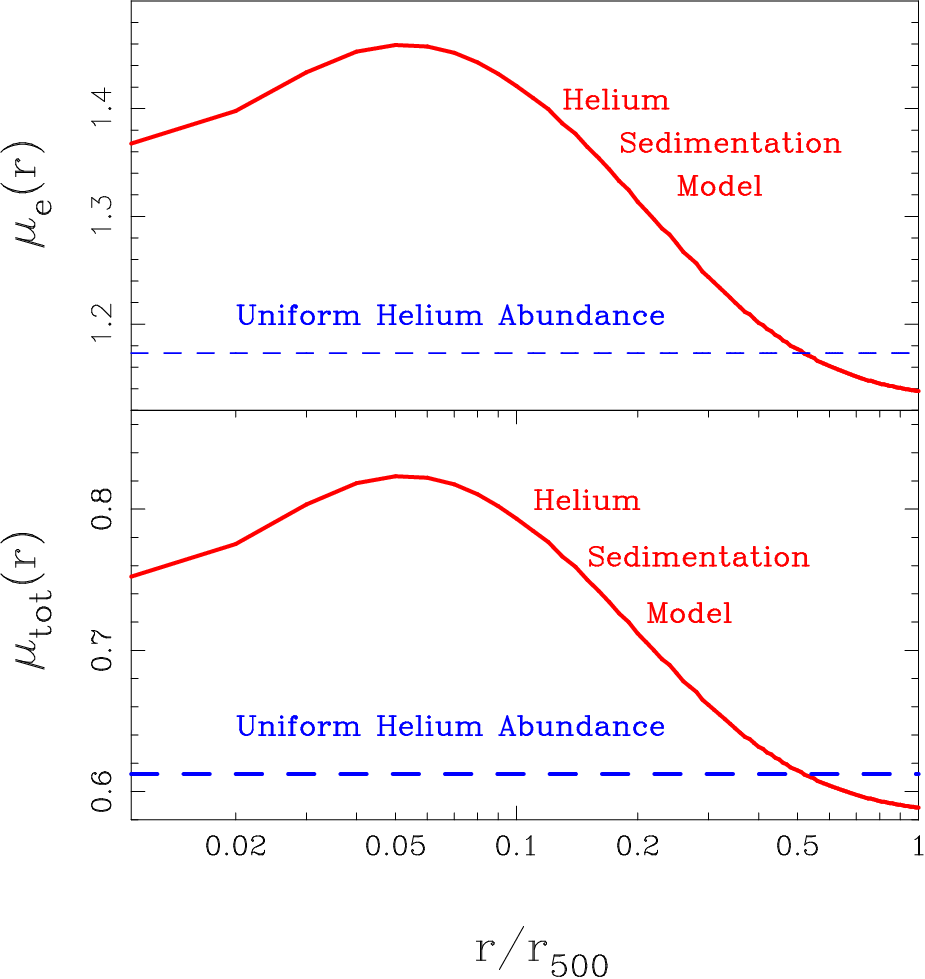}
\hspace{0.03\textwidth}
\caption{The distributions of the electron mean molecular weight (top panel)
and the total mean molecular weight (bottom panel) as a function of radius for a uniform \citet{anders1989} 
\he\ abundance (blue dashed line) and for the 11 Gyr \he\ sedimentation model 
\citet{peng2009}(red line). }
\label{fig:mu}
\vspace{5 mm}
\end{figure}

The gas mass is the volumetric integral of the number density of electrons
multiplied by the electron mean molecular weight:

\begin{equation}
M_{gas}(r)=m_{p}\int_{V} \mu_{e}(r)\ n_{e}(r)\  dV,
\label{eqn:gasMass}
\end{equation}

\noindent where $m_{p}$ is the proton mass and $n_{e}(r)$ is defined 
in Equation \ref{eqn:coolcore_gas_density}. 
Since the mean molecular weight of electrons $\mu_{e}(r)$ is dependent on 
the ion distribution in the plasma, we use the same approach as in the X-ray
cooling function calculations to determine its radial
distribution. The electron mean molecular weight $\mu_{e}(r)$ ,
which is calculated for a uniform \he\
abundance \citep{anders1989} and the radial \he\ abundance profile of  \citep{peng2009} obtained for a sedimentation timescale
of 11 Gyr with no magnetic field,
is shown in the top panel of Figure \ref{fig:mu}.

Similarly, the total mass is the volumetric integral
of the total matter density $\rho_{tot}(r)$,
\begin{equation}
M_{tot}(r)= \int_{V} \ \rho_{tot}(r) \ dV.
\label{eqn:totalMass}
\end{equation}
As in \citet{bulbul2010}, the total matter density is 
found by
\begin{equation}
\rho_{tot}(r)=\left[\frac{T_{0}k(n+1)(\beta-1)}{4\pi G m_{p}r_{s}^{2}}\right]
\frac{1}{\mu_{tot}(r)}\frac{1}{(1+r/r_{s})r/r_{s}}.
\label{eqn:totalDensity}
\end{equation}
where $k$ is the Boltzmann constant, $G$ is Newton's gravitational constant, and 
$\mu_{tot}(r)$ is the total mean molecular weight, which is also dependent on the ion distribution 
in the plasma. The total mean molecular weight, $\mu_{tot}(r)$ is calculated assuming a uniform \he\
distribution and the \he\ sedimentation model (see Figure \ref{fig:mu} bottom panel) as
was done for the $\mu_{e}$ calculations. 
%%%%%%%%%%%%%%%%%%%%%%%%%%%%%%%%%%%%%%%%%%%%%%%%%%%%%%%%%%%%%

\begin{figure*}[htb!]
\centering
\includegraphics[angle=-90,width=9cm]{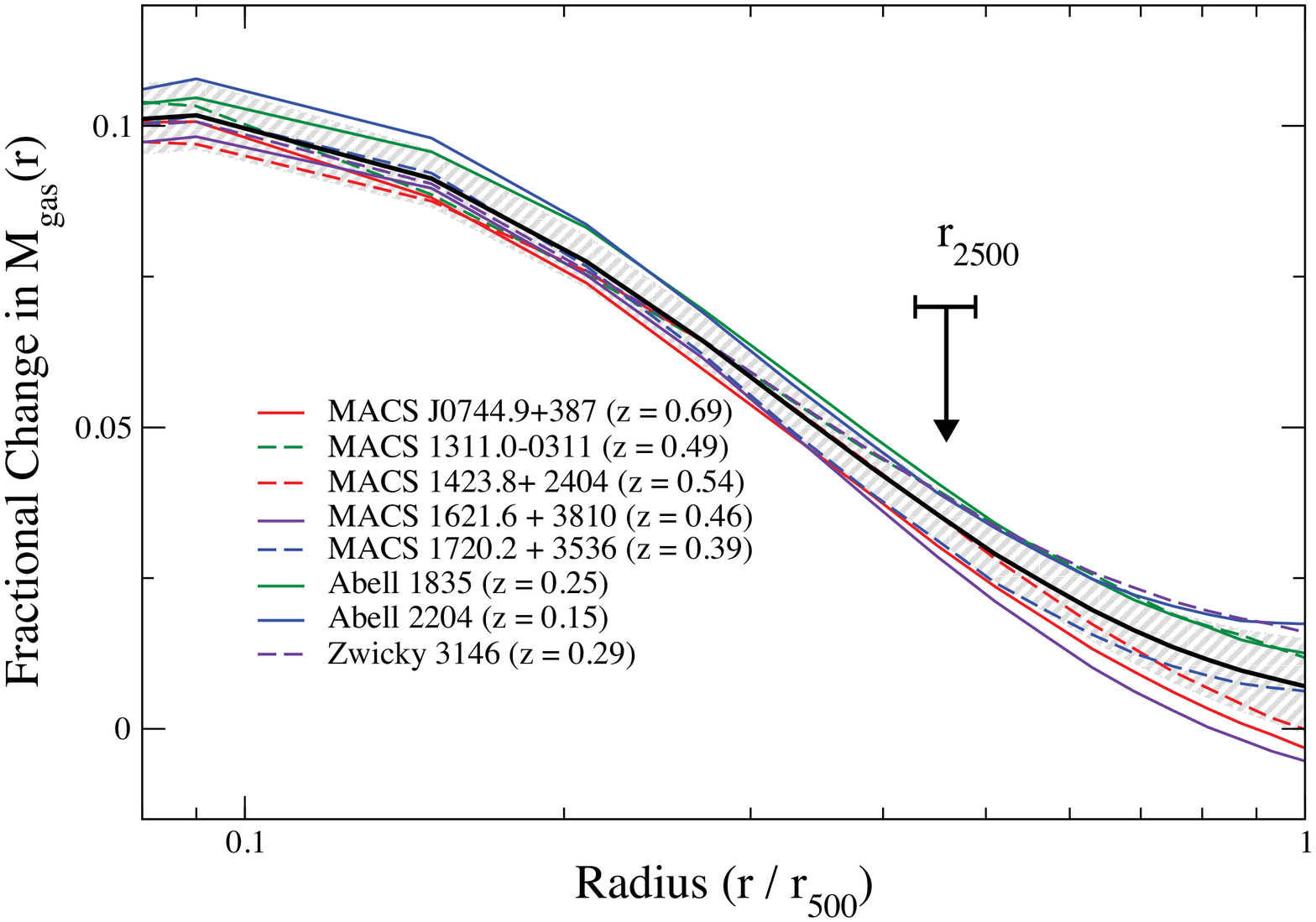}
\hspace{0.01\textwidth}
\includegraphics[angle=-90,width=9cm]{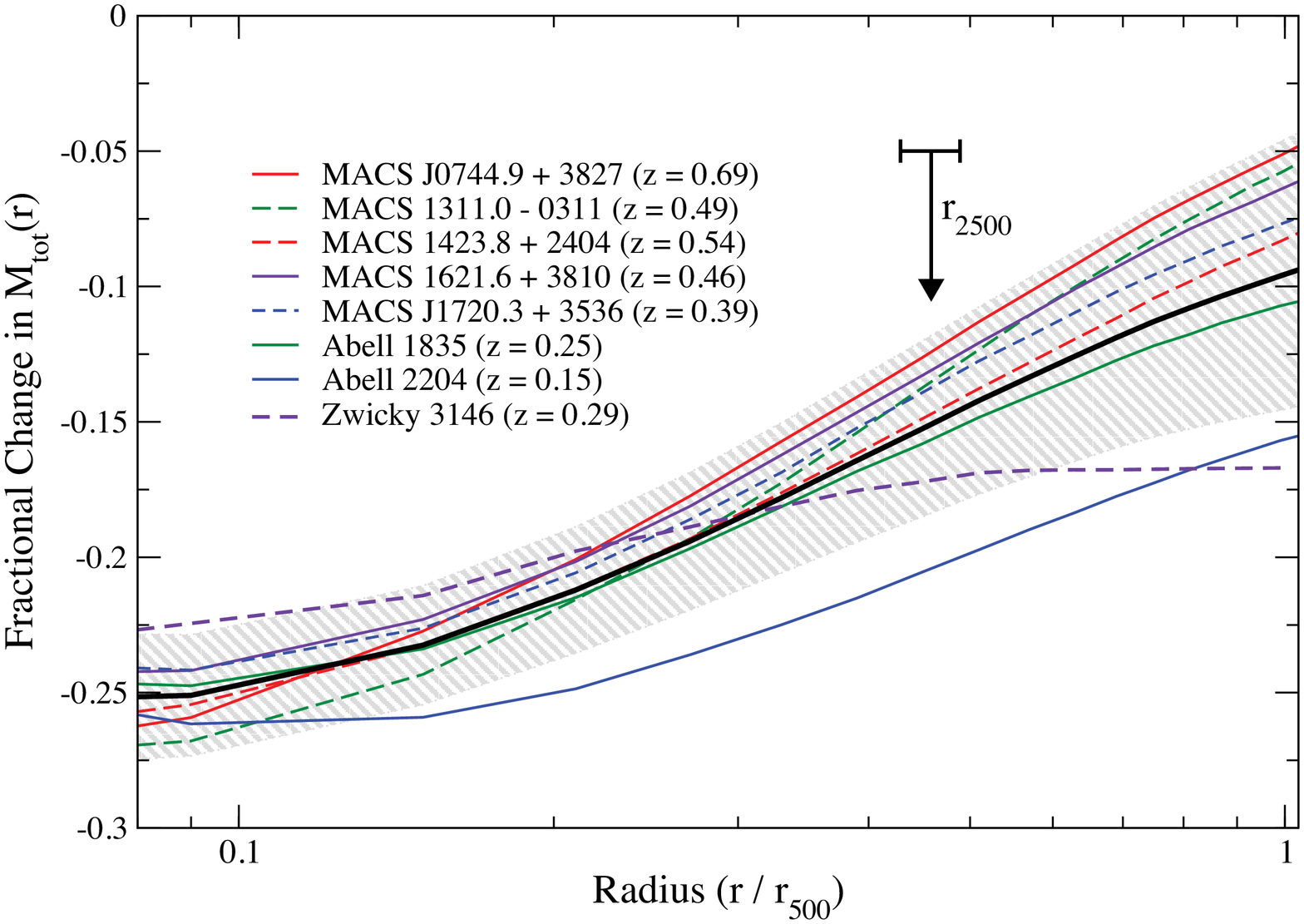}
\caption{The fractional change in gas mass and total mass with radius as a result of an extreme \he~sedimentation with respect to masses produced by a uniform \citet{anders1989} \he\ abundance for a sample of eight relaxed clusters. The black lines correspond to the mean fractional change and the shaded are indicates the standard 
deviation of the fractional change in gas mass and total mass measurements.	
}
\label{fig:massFraction}
\end{figure*}

\section{Results}
\label{section:heliumSedimentation}

We measure cluster masses and scaling relations assuming the limiting cases of uniform
\he~abundance \citep{anders1989} and the radially-varying, sedimented \he~abundance profile of \cite{peng2009}
described in Section~\ref{sec:emissivity}.

%\subsection{Cluster Masses}
%\label{sec:HeSedMasses}

In order to determine the fractional change in cluster mass measurements introduced by the spatial variation in the \he~distribution, we calculate gas mass and total mass for each \he\ abundance model. 
Gas mass is calculated using Equation \ref{eqn:gasMass} and the electron mean molecular weight
distributions shown in the top panel of Figure \ref{fig:mu}. The total mass is calculated using Equation \ref{eqn:totalMass} and the total mean molecular weight distributions shown in the bottom panel of Figure \ref{fig:mu}.
The gas mass and total mass at $r_{2500}$ and $r_{500}$ are given in Tables \ref{table:masses_A&G}
and \ref{table:masses_P&N} for both the uniform \he~abundance and the sedimented case.

\begin{table*}[b!]
\begin{center}
\scriptsize
\caption{Best-fit Parameters of the \citet{bulbul2010} Model Produced Using a Uniform \he\ Abundance Model \citep{anders1989}}
\vspace{1 mm}
\begin{tabular}{lcccccccccc}
\hline

Cluster         & $n_{e0}$              & $r_{s}$             & $n$  & $\beta$       
&$T_{0}$         & $r_{cool}$            &$\alpha$              & $\gamma$        & $\chi^{2}$ (d.o.f.) & \textit{P} value  \\
                & ($10^{-2} cm ^{-3}$)& (\textit{arcsec}) &      & &
($keV$)       & (\textit{arcsec})       &                       &               &  & \\
\\
\hline\hline
\\
MACS~J0744.9+3927& 2.13$^{+0.69}_{-0.30}$ 	& 13.64$^{+2.14}_{-3.18}$ 	& 2.77$^{+0.15}_{-0.18}$ 
& 2.0						& 22.47$^{+3.73}_{-1.39}$	& 15.99$^{+4.56}_{-3.82}$ 		
& 0.13$^{+0.08}_{-0.02}$ 			& 1.0				& 44.3 (51)	& 73.5\%	\\
\\
Zwicky 3146     & 1.27$^{+0.43}_{-0.06}$ 	& 52.99$^{+5.47}_{-5.43}$	& 2.58$^{+0.83}_{-0.77}$
& 2.85$^{+0.83}_{-0.18}$ 			& 30.02$^{+1.99}_{-2.31}$	& 58.95$^{+5.99}_{-6.07}$
& 0.06$^{+0.02}_{-0.02}$			& 1.0				& 66.8 (78)	& 81.2\% \\
\\ 
MACS~J1311.0-0311& 1.73$^{+0.20}_{-0.65}$	& 40.32$^{+5.69}_{-6.31}$	& 5.38$^{+3.19}_{-0.47}$		
& 2.0						& 11.36$^{+1.46}_{-0.71}$	& 20.73$^{+4.36}_{-1.36}$		
& 0.33$^{+0.04}_{-0.13}$			& 2.0				& 70.56 (63)	& 42.5 \%	\\			
\\ 
Abell 1835      & 2.57$^{+0.29}_{-0.07}$	 & 40.32$^{+3.29}_{-6.48}$ 	& 3.98$^{+0.73}_{-0.41}$
& 1.94$^{+0.15}_{-0.22}$                	 & 18.26$^{+0.42}_{-1.61}$ 	& 22.65$^{+0.28}_{-1.17}$  
& 0.18$^{+0.02}_{-0.01}$			 & 2.0	       			& 99.3 (93)     & 30.8\%        \\
\\
MACS~J1423.8+2404&3.29$^{+0.68}_{-0.29}$	& 11.39$^{+1.60}_{-2.17}$	& 2.98$^{+0.14}_{-0.16}$
& 2.0						& 16.03$^{+1.00}_{-1.32}$	& 13.07$^{+0.63}_{-0.73}$
& 0.18$^{+0.04}_{-0.02}$			& 2.0				& 45.5 (50)	& 65.3\%	\\
\\
MACS~J1621.6+3810& 2.84$^{+0.62}_{-0.16}$	& 14.39$^{+3.85}_{-2.85}$	& 3.12$^{+0.17}_{-0.12}$
& 2.0		 				& 12.63$^{+0.95}_{-0.97}$	& 9.01$^{+0.73}_{-0.86}$\
& 0.26$^{+0.05}_{-0.03}$			& 2.0				& 32.4 (45)	& 92.1\%	\\
\\ 
Abell 2204      & 4.42$^{+0.37}_{-0.24}$ 	& 21.73$^{+1.50}_{-2.01}$ 	& 6.44$^{+1.02}_{-0.51}$
& 1.39$^{+0.04}_{-0.06}$	                & 14.28$^{+0.75}_{-0.78}$	& 19.42$^{+0.60}_{-0.73}$  
& 0.16$^{+0.01}_{-0.01}$			& 2.0          			& 115.5 (145)  & 96.6\%       \\
\\
MACS~J1720.3+3536&2.12$^{+0.24}_{-0.20}$	& 31.69$^{+5.58}_{-5.03}$	& 3.97$^{+0.29}_{-0.27}$	
& 2.0						& 13.01$^{+0.97}_{-0.79}$	& 10.95$^{+0.86}_{-0.80}$
& 0.13$^{+0.02}_{-0.01}$			& 2.0				& 46.5 (60)	& 89.9\%	\\
\\
\hline 
\label{table:bestFitParams_A&G}
\end{tabular}
\end{center}
\end{table*}

\begin{table*}
\begin{center}
\scriptsize
\caption{Best-fit Parameters of the \citet{bulbul2010} Produced Using a \he\ Sedimentation Model \citep{peng2009}}
\vspace{1 mm}
\begin{tabular}{lcccccccccc}
\hline

Cluster         & $n_{e0}$             		 & $r_{s}$             		& $n$  
& $\beta$       				 & $T_{0}$         		& $r_{cool}$
& $\alpha$ 				         & $\gamma$        		& $\chi^{2}$ (d.o.f.) & \textit{P} value  \\
                & ($10^{-2} cm ^{-3}$)		 & (\textit{arcsec}) 		&     
&  						 & ($keV$)       		& (\textit{arcsec})
& 				                 &            			&  & \\
\\
\hline\hline
\\  
MACS~J0744.9+3927& 2.07$^{+0.26}_{-0.25}$ 	& 14.79$^{+1.37}_{-1.88}$ 	& 2.84$^{+0.10}_{-0.15}$ 
& 2.0						& 21.75$^{+1.71}_{-1.91}$	& 12.19$^{+3.51}_{-2.24}$ 		
& 0.13 						& 1.0				& 45.7 (52)	& 71.85\%	\\
\\
Zwicky 3146     &1.28$^{+0.10}_{-0.07}$		& 51.37$^{+7.53}_{-5.16}$	& 2.68$^{+0.39}_{-0.24}$	
& 2.68$^{+0.37}_{-0.13}$ 			& 25.92$^{+1.88}_{-0.91}$	& 47.33$^{+3.63}_{-5.62}$	
& 0.11						& 1.0				& 60.83 (79)	& 93.6\% \\
\\
MACS~J1311.0-0311& 1.56$^{+0.07}_{-0.09}$	& 42.26$^{+6.41}_{-7.56}$	& 5.38$^{+0.54}_{-0.67}$
& 2.0						& 10.91$^{+1.01}_{-0.76}$	& 19.32$^{+1.94}_{-1.16}$
& 0.34						& 2.0				& 44.6 (64)	& 68.9\%	\\
\\
Abell 1835      & 2.36$^{+0.30}_{-0.05}$	& 38.75$^{+5.10}_{-1.18}$	& 4.01$^{+0.28}_{-0.83}$	
& 2.88$^{+0.35}_{-0.29}$			& 17.75$^{+0.91}_{-0.16}$	& 21.43$^{+0.42}_{-0.39}$	
& 0.18						& 2.0				& 108.1 (94)	& 13.6 \%		\\
\\
MACS~J1423.8+2404& 2.99$^{+0.11}_{-0.09}$	& 11.73$^{+0.86}_{-0.84}$	& 2.93$^{+0.11}_{-0.54}$
& 2.0						& 15.49$^{+0.98}_{-1.28}$	& 12.28$^{+0.45}_{-0.54}$
& 0.18						& 2.0				& 47.2 (51)	&  58.5 \%	\\
\\
MACS~J1621.6+3810& 2.58$^{+0.16}_{-0.14}$	& 14.61$^{+1.10}_{-1.26}$	& 3.06$^{+0.09}_{-0.12}$
& 2.0						& 12.66$^{+0.95}_{-1.18}$	& 8.29$^{+0.69}_{-0.62}$
& 0.26						& 2.0				& 33.88 (46)	& 88.8 \%		\\
\\
Abell 2204      &4.15$^{+0.07}_{-0.08}$		& 19.81$^{+0.43}_{-0.56}$	& 5.59$^{+0.66}_{-0.39}$	
& 2.43$^{+0.04}_{-0.05}$			& 14.51$^{+0.53}_{-0.59}$	& 18.81$^{+0.39}_{-0.28}$	
& 0.16						& 2.0				& 127.7 (146)	& 84.6 \% \\
\\
MACS~J1720.3+3536&1.91$^{+0.13}_{-0.12}$	& 31.62$^{+3.71}_{-5.30}$	& 3.85$^{+0.21}_{-0.19}$	
& 2.0						& 12.96$^{+1.01}_{-0.79}$	& 10.57$^{+0.66}_{-0.63}$
& 0.13						& 2.0				& 47.3(61)	& 90.0\%	\\
\\ 
\hline
\label{table:bestFitParams_P&N}
\end{tabular}
\end{center}
\end{table*}

\begin{table*}
\centering
\caption{Gas Mass and Total Mass  for the Uniform \he\ Abundance Model  at the Overdensity Radii 
$r_{2500}$ and $r_{500}$}
\vspace{1 mm}
\begin{tabular}{@{\extracolsep{\fill}}lccccccc}
\hline\hline

Cluster         & $r_{2500}$            & $M_{gas}(r_{2500})$    	& $M_{tot}(r_{2500})$   
		  & $r_{500}$     		& $M_{gas}(r_{500})$ 		& $M_{tot}(r_{500})$ 	\\
		    &(arcsec)              	 & ($10^{13} M_{\odot}$) 	& ($10^{14} M_{\odot}$) 
		& (arcsec)      		& ($10^{13} M_{\odot}$) 	& ($10^{14} M_{\odot}$)  \\
\hline
\\
MACS~J0744.9+3927& 57.9$^{+3.1}_{-2.8}$ 	& 2.98$^{+0.23}_{-0.20}$	& 2.07$^{+0.36}_{-0.29}$	
		 &120.2$^{+8.1}_{-6.8}$ 	& 8.78$^{+0.66}_{-0.55}$	& 3.69$^{+0.80}_{-0.59}$\\
\\	
Zwicky 3146     & 127.6$^{+4.5}_{-5.3}$ 	& 4.26$^{+0.18}_{-0.20}$	& 3.17$^{+0.35}_{-0.38}$  
		& 266.7$^{+12.4}_{-14.2}$	& 9.72$^{+0.39}_{-0.44}$	& 5.78$^{+1.23}_{-1.10}$ \\
\\
MACS~J1311.0-0311& 78.7$^{+5.9}_{-5.5}$ 	& 2.29$^{+0.19}_{-0.18}$	& 2.51$^{+0.61}_{-0.49}$ 
		 & 173.1$^{+17.1}_{-15.1}$	& 5.16$^{+0.29}_{-0.29}$	& 5.34$^{+1.74}_{-1.28}$ \\
\\
Abell~1835      & 150.6$^{+3.4}_{-4.2}$ 	& 4.97$^{+0.14}_{-0.17}$ 	& 3.72$^{+0.26}_{-0.30}$

                & 309.7$^{+9.8}_{-13.1}$ 	& 12.08$^{+0.38}_{-0.50}$	& 6.38$^{+0.64}_{-0.79}$\\
\\
MACS~J1423.8+2404& 63.4$^{+2.4}_{-2.8}$ 	& 2.16$^{+0.10}_{-0.11}$	& 1.61$^{+0.19}_{-0.21}$
		 & 125.9$^{+5.4}_{-6.4}$	& 5.46$^{+0.23}_{-0.27}$	& 2.52$^{+0.34}_{-0.37}$\\
\\
MACS~J1621.6+3810& 67.8$^{+4.2}_{-3.5}$		& 1.78$^{+0.15}_{-0.12}$	& 1.37$^{+0.27}_{-0.20}$  
		 & 135.1$^{+9.3}_{-7.8}$	& 4.77$^{+0.31}_{-0.26}$	& 2.17$^{+0.48}_{-0.35}$ \\
\\
Abell~2204      & 225.7$^{+4.1}_{-4.1}$ 	& 3.99$^{+0.09}_{-0.09}$ 	& 3.37$^{+0.19}_{-0.18}$

                & 479.8$^{+11.4}_{-11.2}$ 	& 10.35$^{+0.26}_{-0.26}$	& 6.48$^{+0.47}_{-0.44}$\\
\\
MACS~J1720.3+3536& 91.6$^{+4.2}_{-3.9}$  	& 2.82$^{+0.18}_{-0.16}$   	& 2.34$^{+0.34}_{-0.28}$
                 & 190.5$^{+10.6}_{-9.3}$	& 7.38$^{+0.35}_{-0.33}$	& 4.20$^{+0.74}_{-0.59}$\\
\\	
\hline

\label{table:masses_A&G}
\end{tabular}
\end{table*}

\begin{table*}
\centering
\caption{Gas Mass and Total Mass for the \he\ Sedimentation Model  at the Overdensity Radii 
$r_{2500}$ and $r_{500}$ }
\vspace{1 mm}
\begin{tabular}{@{\extracolsep{\fill}}lccccccc}
\hline
\\
Cluster         & $r_{2500}$            & $M_{gas}(r_{2500})$    	& $M_{tot}(r_{2500})$   
		& $r_{500}$     	& $M_{gas}(r_{500})$ 		& $M_{tot}(r_{500})$ 	\\
                &(arcsec)               & ($10^{13} M_{\odot}$) 	& ($10^{14} M_{\odot}$) 
		& (arcsec)      	& ($10^{13} M_{\odot}$) 	& ($10^{14} M_{\odot}$)  \\
\hline
\\
MACS~J0744.9+3927& 53.9$^{+2.7}_{-3.1}$	& 2.75$^{+0.21}_{-0.22}$	& 1.66$^{+0.27}_{-0.27}$
                & 116.9$^{+6.0}_{-6.9}$	& 8.43$^{+0.50}_{-0.54}$	& 3.41$^{+0.56}_{-0.56}$\\
\\
Zwicky 3146     & 116.0$^{+4.2}_{-3.7}$ & 3.94$^{+0.17}_{-0.15}$	& 2.38$^{+0.27}_{-0.22}$
                & 246.5$^{+13.1}_{-11.5}$ & 9.14$^{+0.44}_{-0.38}$	& 4.56$^{+0.77}_{-0.61}$\\
\\
MACS~J1311.0-0311& 71.2$^{+3.9}_{-4.3}$	& 2.12$^{+0.12}_{-0.15}$	& 1.86$^{+0.32}_{-0.32}$
		 & 165.6$^{+9.2}_{-10.5}$& 5.06$^{+0.16}_{-0.19}$	& 4.67$^{+0.82}_{-0.83}$\\
\\
Abell~1835      & 138.6$^{+2.6}_{-4.3}$ & 4.62$^{+0.10}_{-0.17}$	& 2.89$^{+0.16}_{-0.26}$
                & 294.8$^{+7.7}_{-12.7}$& 11.59$^{+0.30}_{-0.51}$	& 5.58$^{+0.45}_{-0.69}$\\
\\
MACS~J1423.8+2404& 58.8$^{+2.4}_{-2.2}$& 2.01$^{+0.10}_{-0.01}$		& 1.29$^{+0.17}_{-0.14}$
		 & 120.9$^{+4.8}_{-4.2}$& 5.20$^{+0.19}_{-0.18}$	& 2.23$^{+0.27}_{-0.23}$\\
\\
MACS~J1621.6+3810& 63.4$^{+3.4}_{-3.4}$	& 1.65$^{+0.11}_{-0.11}$	& 1.12$^{+0.19}_{-0.17}$
		 & 130.7$^{+6.9}_{-6.8}$& 4.55$^{+0.23}_{-0.23}$	& 1.97$^{+0.33}_{-0.29}$\\
\\
Abell~2204	& 201.0$^{+3.6}_{-3.9}$& 3.55$^{+0.08}_{-0.09}$		& 2.38$^{+0.13}_{-0.14}$
                & 440.4$^{+11.5}_{-10.9}$& 9.53$^{+0.28}_{-0.25}$       & 5.01$^{+0.40}_{-0.36}$\\
\\

MACS~J1720.3+3536& 84.6$^{+3.9}_{-3.5}$	& 2.59$^{+0.16}_{-0.15}$	& 1.84$^{+0.27}_{-0.22}$
		 & 182.9$^{+8.5}_{-8.1}$& 7.09$^{+0.31}_{-0.29}$	& 3.72$^{+0.54}_{-0.47}$\\
\\
\hline
\label{table:masses_P&N}
\end{tabular}
\end{table*}

\begin{table*}
\centering
\normalsize
\caption{\label{table:percentages}Percentage Changes in $M_{gas}$ and $M_{tot}$}
\vspace{2 mm}
\begin{tabular}{lcc}
\hline
Region 			&\% Change in M$_{gas}$ &\% Change in M$_{tot}$\\
\hline\\
$0.0<r<0.1r_{500}$	& 10.5 $\pm$ 0.8 	& 25.1 $\pm$ 1.1 \\
\\
$0.0<r<r_{2500}$	& 3.5 $\pm$ 1.0		& 13.9 $\pm$ 3.1\\
\\
$0.0<r<r_{500}$	& 1.4 $\pm$ 1.2		& 12.5 $\pm$ 5.5\\ 
\\
\hline
\end{tabular}
\vspace{5 mm}
\end{table*}

The fractional change with radius in gas mass and total mass measurements
as a result of an extreme \he\ sedimentation case
is shown in Figure \ref{fig:massFraction} for each of the relaxed galaxy clusters in our sample. The black line shows the mean fractional change 
in gas mass (left panel) and total mass (right panel). The shaded areas indicate the standard deviations
of the gas mass and total mass inferred for the sample.
In Table \ref{table:percentages} we show the weighted mean percentage difference 
with rms errors in gas mass and total mass measurements within small cluster radius ($r < 0.1 r_{500}$), 
intermediate radius ($r < r_{2500}$) and large radius ($r < r_{500}$). 
On average, at large radii ($r < r_{500}$) the effect of \he\
sedimentation on gas mass is negligible (1.3 $\pm$ 1.2 \%). The \he\
sedimentation model produces an average of 10.2 $\pm$ 5.5  \% 
decrease in the inferred total mass within $r_{500}$.
Significantly stronger effects in the gas mass
(10.5 $\pm$ 0.8 \%) and 
total mass (25.1 $\pm$ 1.1 \%) are seen at small cluster radii ($r < 0.1 r_{500}$) 
where the \he\ abundance enhancement and gradient are greater. This result is in agreement with 
the expected change in the mass measurements predicted by theoretical studies
\citep{qin2000,peng2009}.

\begin{figure*}[ht!]
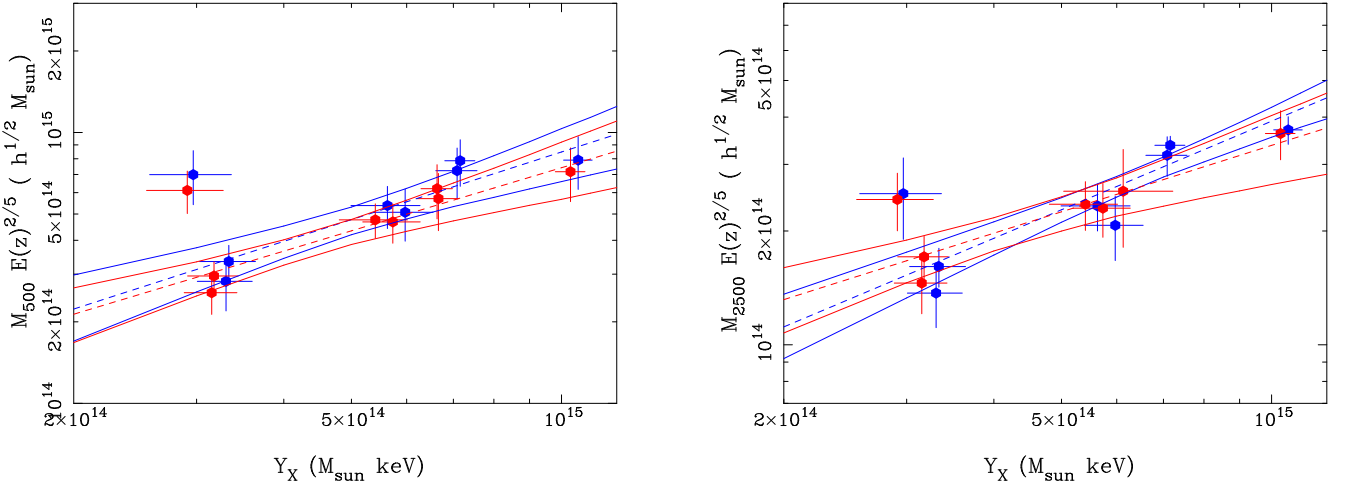

\centering
\includegraphics[angle=-90,width=0.44\textwidth]{f6a_new.eps}
\hspace{0.03\textwidth}
\hspace{0.02\textwidth}
\includegraphics[angle=-90,width=0.44\textwidth]{f6b_new.eps}
\caption{The Y$_{X}$-M$_{500}$ and Y$_{X}$-M$_{2500}$ scaling relations for a sample of
eight relaxed galaxy clusters. The blue data points and curves correspond to the measurements obtained from a uniform \he\ abundance case and red data points and curves correspond to the measurements obtained from
a \he\ sedimentation model. The dashed lines show the best fit  
power law relation and solid curves show the 90\% confidence levels.
}
\label{fig:YxVsMtot}
\vspace{5 mm}
\end{figure*}

\subsection{$Y_{X}$-$M$ Scaling Relation}

We examine the effect of \he\ sedimentation on the cluster mass scaling relations.
The X-ray mass proxy, $Y_{X}$, is defined as \citep{kravtsov2006},

\begin{equation}
Y_{X}=M_{gas}(r_{500})\ T_{X}.
\end{equation}

\noindent where $T_{X}$ is the mean X-ray spectroscopic temperature 
measured within $r_{500}$ \citep{kravtsov2006,vikhlinin2009}.
The importance of this quantity is that there is a 
correlation between $Y_{X}$ and the total cluster mass,
$M_{500}$ \citep{kravtsov2006,vikhlinin2009}
and this correlation motivates $Y_{X}$ as a
proxy for total mass \citep{vikhlinin2009,andersson2010}. In this section we investigate the effect of \he\
sedimentation on the $Y_{X}$-$M_\Delta$ scaling relation 
at overdensities $\Delta=(2500,500)$ (i.e. for properties computed within $r < (r_{2500}, r_{500})$).

The $Y_{X}$-$M_{500}$ scaling relation is found using
gas mass and total mass measurements shown in 
Tables \ref{table:masses_A&G} and \ref{table:masses_P&N} 
for a uniform \he\ abundance and \he\ sedimentation distributions.
The average spectral temperature $T_{X}$ is obtained with a
single-temperature thermal plasma model (\apec) fit to 
the \chandra\ spectrum in the radial range $0.15r_{500}<r<r_{500}$.
Using the XSPEC analysis package to vary the \he\ abundance assumed in our spectroscopic fits,
we determine that the assumed \he\ distribution
has a negligible effect ($\lesssim$ 1\%) on the measurements of the
average X-ray spectral temperatures $T_{X}$. 
We report the average X-ray temperature $T_{X}$ and the X-ray mass proxy $Y_{X}$
for a uniform \he\ distribution and \he\ sedimentation model
in Table \ref{table:temperature}.

\begin{table*}
\centering
\footnotesize
\caption{\label{table:temperature}Measured Average Temperature ($T_{X}$) and the Mass Proxy ($Y_{X}$)}
\vspace{2 mm}
\begin{tabular}{lccc}
\hline
Cluster			&$T_{X}$	&$Y_{X}$&$Y_{X}$\\
			&		&(Uniform \he Abundance)&( \he\ Sedimentation)\\
			&(keV)		& ($10^{14} keV M_{\odot}$)&($10^{14} keV M_{\odot}$)\\
\hline\\
MACS~J0744.9+3927	& 6.80 $\pm$ 0.5& 5.97 $\pm$ 0.58& 5.73 $\pm$ 0.54\\
\\
Zwicky~3146		& 7.29 $\pm$ 0.4& 7.09 $\pm$ 0.48& 6.66 $\pm$ 0.45\\
\\
MACS~J1311.0-0311	& 5.75 $\pm$ 0.7& 2.97 $\pm$ 0.39& 2.91 $\pm$ 0.37\\	
\\
Abell~1835		& 8.79 $\pm$ 0.3& 10.57 $\pm$ 0.49& 10.30 $\pm$ 0.50\\	
\\
MACS~J1423.8+2404	& 6.11 $\pm$ 0.5& 3.34 $\pm$ 0.31& 3.18 $\pm$ 0.27\\	
\\
MACS~J1621.6+3810	& 6.93 $\pm$ 0.5& 3.31 $\pm$ 0.29& 3.15 $\pm$ 0.27 \\	
\\
Abell~2204 		& 6.91 $\pm$ 0.3& 7.16 $\pm$ 0.36& 6.63 $\pm$ 0.34 \\	
\\
MACS~J1720.3+3536	& 7.63 $\pm$ 0.8& 5.63 $\pm$ 0.64& 5.41 $\pm$ 0.61\\	
\\
\hline
\end{tabular}
\vspace{5 mm}
\end{table*}

The $Y_{X}$-$M_{500}$ scaling relation for our sample of eight relaxed galaxy clusters
is shown in Figure \ref{fig:YxVsMtot} (left panel).
For each $\Delta$, we fit the $Y_{X}$-$M_\Delta$ data using a power 
law relation \citep{vikhlinin2009},

\begin{equation}
\begin{aligned}
M_\Delta \ E(z)^{2/5}&=a \left( \frac{Y_{X}}{3 \times10^{14}M_{\odot}keV}\right)^b  \ h^{1/2} M_{\odot} ,
\end{aligned}
\label{eqn:powerLawRelation}
\end{equation}

\noindent where $a$ is the normalization, $b$ is the slope, and  $E(z)$ is the evolution 
function. The best fit normalizations and slopes are reported in Table \ref{table:slopes}
with the goodness of the fit. In Figure \ref{fig:YxVsMtot} 
the dashed lines shows the best fit power law model and solid lines 
shows the 90 \% confidence intervals for both a uniform \he\ abundance and the
\he\ sedimentation cases.

From the $Y_{X}$-$M_{500}$ fit to the data on the eight galaxy clusters in our sample, we conclude that 
both the uniform and sedimented \he\ fits are statistically
consistent with the power law slope found by \citet{vikhlinin2009}
(see Table \ref{table:slopes}). \he\ sedimentation changes the normalization by 10 $\pm$ 0.9 \%
at $r_{500}$. 
The difference in best-fit slope between the uniform and \he\
sedimentation models is small compared to the statistical uncertainty.

We also investigate the effect of \he\ sedimentation on the
Y$_{X}$-$M_{2500}$ scaling relation.
Figure \ref{fig:YxVsMtot} (right panel) shows the $Y_{X}$-$M_{2500}$ 
scaling relation for our sample.
We fit the $Y_{X}$-$M_{2500}$ scaling relation data with a power law 
shown in Equation \ref{eqn:powerLawRelation}.  
The dashed lines show the best fit power law relation
with the goodness of fit reported in Table \ref{table:slopes_r2500}
for the uniform and sedimented \he\ profiles.
In Figure \ref{fig:YxVsMtot} the best fit 
power law model is shown in dashed line and 90\% confidence 
levels are shown in solid lines for both uniform \he\
abundance and \he\ sedimentation models.

The self-similar $Y_{X}-M_{2500}$ scaling relation discussed 
by \citet{bonamente2008} is of the form:

\begin{equation}
Y_{X}\propto M_{2500}^{5/3} E(z)^{2/3}
\end{equation}

which is equivalent to the $Y_{X}-M_{2500}$ scaling relation 
form we use for this work 

\begin{equation}
M_{2500} E(z)^{2/5} \propto Y_{X}^{3/5}.
\label{eqn:selfSimilarValue}
\end{equation}

Recently, \citet{bonamente2008} measured the $Y_{X}-M_{2500}$ 
scaling relation, and found it to be consistent with the
self-similar prediction, $b=1.66\pm0.20$ in Equation \ref{eqn:powerLawRelation}.
In our fit $Y_{X}-M_{2500}$ (shown in Table \ref{table:slopes_r2500}), we also find that
both the uniform \he\ abundance and the \he\ sedimentation
cases are consistent with the slope measured by \citet{bonamente2008}
and with the self-similar expectation (Equation \ref{eqn:selfSimilarValue}).
The difference in best-fit slopes $b$ is negligible compared to the 
statistical uncertainty reported in Table \ref{table:slopes_r2500}.
However, we find that \he\ sedimentation affects the normalization of 
the $Y_{X}-M_{2500}$ scaling relation by 13 $\pm$ 0.6 \%.

\begin{table*}
\centering
\footnotesize
\caption{\label{table:slopes}Comparison of Normalization and Slope of the
 $Y_{X}-M_{500}$ Scaling Relation}
\vspace{2 mm}
\begin{tabular}{lcccc}
\hline
					&a		 & b	& $\chi^{2}$ (d.o.f)& P (\%)\\
					&($10^{14}\ h^{1/2}\ M_{\odot}$)&\\
\hline\\
Uniform \he\ Abundance 	& 3.12 $\pm$ 0.87 & 0.83 $\pm$ 0.42	& 8.6 (6) & 19.7\\
\\
\he\ Sedimentation Model & 2.92$\pm$ 0.61  & 0.77 $\pm$ 0.22	& 10.7 (6) & 9.8\\
\\
\hline
\end{tabular}
\vspace{5 mm}
\end{table*}

\begin{table*}
\centering
\footnotesize
\caption{\label{table:slopes_r2500}Comparison of Normalization and Slope 
of the $Y_{X}-M_{2500}$ Scaling Relation}
\vspace{1 mm}
\begin{tabular}{lcccc}
\hline
					&a		&b		& $\chi^{2}$(d.o.f.) & P\\
			& ($10^{14}\ h^{1/2}\ M_{\odot}$)&\\
\hline\\

Uniform \he\ Abundance 	& 1.94 $\pm$ 0.50 & 0.72 $\pm$ 0.28& 5.0 (6)& 54.3\\
\\
\he\ Sedimentation Model & 1.66 $\pm$ 0.30 & 0.66 $\pm$ 0.11& 5.6 (6)& 46.9 \\
\hline
\end{tabular}
\vspace{5 mm}
\end{table*}

\section{Conclusion }
\label{section:conclusion}

In this paper we investigate the upper limits to the effect of 
the \he\ sedimentation on X-ray derived cluster masses and the $Y_{X}-M_{500}$ and
$Y_{X}-M_{2500}$ scaling relations using \chandra\ X-ray observations of eight relaxed galaxy clusters. 
We used a limiting \he\ sedimentation profile for 11 Gyr old clusters based on the 
simulations performed by \citet{peng2009} and compare these results with those assuming 
a uniform \he\ distribution in order to determine the maximum impact of \he\ sedimentation.
This work is the first application of a theoretically
predicted \he\ sedimentation profile to \chandra\ observations.
We fit the deep exposure \chandra\ X-ray spectroscopic and imaging data of the
eight relaxed galaxy clusters with an analytic model of the intra-cluster plasma
developed by \citet{bulbul2010}. We demonstrated that both a uniform \he\ and a
limiting \he\ sedimentation
model can accurately describe the surface brightness and temperature profiles obtained from
the \chandra\ X-ray data.

We have found that, on average, the effect of \he\ sedimentation on gas mass inferred 
within large radii ($r< r_{500}$) is negligible (1.3 $\pm$ 1.2 \%). 
The \he\ sedimentation model produces a 10.2 $\pm$ 5.5  \% mean
decrease in the total mass inferred within $r< r_{500}$.
Significantly stronger effects in the gas mass (10.5 $\pm$ 0.8 \%) and 
total mass (25.1 $\pm$ 1.1 \%) are seen at small radii 
owing to a larger variance in \he\ abundance in the inner region, $r \leq0.1\, r_{500}$.
This study supports the view that \he\ sedimentation should 
have a negligible impact on cluster mass inferred within large radii. This result is consistent with
the predictions of the previous theoretical studies by \citet{ettori2006} and \citet{peng2009}.

The fractional change in both the gas mass and the total mass measurements due to helium sedimentation do not show any trend with redshift. The strongest effect on the cluster total mass is observed on an intermediate redshift cluster, Zwicky 3146, while the smallest effect is observed on the highest redshift cluster in the sample, MACS J0744.9+3827.

In order to investigate the effect of \he\ sedimentation
on the best-fit slopes and normalizations of the $Y_{X}-M_{500}$ and $Y_{X}-M_{2500}$ scaling relations,
we used gas mass measurements, and the mean X-ray spectroscopic temperature $T_{X}$ 
to estimate the X-ray mass proxy $Y_{X}$ for both the uniform and sedimented \he\ abundance profiles.
Both uniform and \he\ sedimentation models produce slopes which are statistically
consistent with the power law slopes found by \citet{bonamente2008} and \citet{vikhlinin2009}.
We have found that \he\ sedimentation has a negligible
effect on the slopes of $Y_{X}-M_{500}$ and $Y_{X}-M_{2500}$ scaling relations.
We have also found that \he\ sedimentation changes
the normalization of the $Y_{X}-M_{500}$ scaling relation 
by 10 $\pm$ 0.9 \% and the $Y_{X}-M_{2500}$ scaling relation
by 13 $\pm$ 0.6 \%.

\section*{Acknowledgments} The authors would like to thank 
Daisuke Nagai and the anonymous referee for comments to improve the manuscript.

\bibliographystyle{aa}

\begin{thebibliography}{34}
\expandafter\ifx\csname natexlab\endcsname\relax\def\natexlab#1{#1}\fi

\bibitem[Abramopoulos et al.(1981)]{abramopoulos1981} Abramopoulos, F., 
Chanan, G.~A., \& Ku, W.~H.-M.\ 1981, \apj, 248, 429 



\bibitem[{{Allen} {et~al.}(2008){Allen}, {Rapetti}, {Schmidt}, {Ebeling},
  {Morris}, \& {Fabian}}]{allen2008}
{Allen}, S.~W., {Rapetti}, D.~A., {Schmidt}, R.~W., {Ebeling}, H., {Morris},
  R.~G., \& {Fabian}, A.~C. 2008, \mnras, 383, 879

\bibitem[Anders 
\& Grevesse(1989)]{anders1989} Anders, E., \& Grevesse, N.\ 1989, \gca, 53, 197 



\bibitem[Andersson et al.(2010)]{andersson2010} Andersson, K., et 
al.\ 2010, arXiv:1006.3068 


\bibitem[{{Birkinshaw} {et~al.}(1991){Birkinshaw}, {Hughes}, \&
  {Arnaud}}]{birkinshaw1991}
{Birkinshaw}, M., {Hughes}, J.~P., \& {Arnaud}, K.~A. 1991, \apj, 379, 466



\bibitem[{{Bonamente} {et~al.}(2004){Bonamente}, {Joy}, {Carlstrom}, {Reese},
  \& {LaRoque}}]{bonamente2004}
{Bonamente}, M., {Joy}, M.~K., {Carlstrom}, J.~E., {Reese}, E.~D., \&
  {LaRoque}, S.~J. 2004, \apj, 614, 56


\bibitem[{{Bonamente} {et~al.}(2006){Bonamente}, {Joy}, {LaRoque}, {Carlstrom},
  {Reese}, \& {Dawson}}]{bonamente2006}
{Bonamente}, M., {Joy}, M.~K., {LaRoque}, S.~J., {Carlstrom}, J.~E., {Reese},
  E.~D., \& {Dawson}, K.~S. 2006, \apj, 647, 25


\bibitem[{{Bonamente} {et~al.}(2008){Bonamente}, {Joy}, {LaRoque}, {Carlstrom},
  {Nagai}, \& {Marrone}}]{bonamente2008}
{Bonamente}, M., {Joy}, M., {LaRoque}, S.~J., {Carlstrom}, J.~E., {Nagai}, D.,
  \& {Marrone}, D.~P. 2008, \apj, 675, 106


\bibitem[Bulbul et al.(2010)]{bulbul2010} Bulbul, G.~E., Hasler, 
N., Bonamente, M., \& Joy, M.\ 2010, \apj, 720, 1038 





\bibitem[Chuzhoy 
\& Nusser(2003)]{chuzhoy2003} Chuzhoy, L., \& Nusser, A.\ 2003, \mnras, 342, L5 


\bibitem[Chuzhoy 
\& Loeb(2004)]{chuzhoy2004} Chuzhoy, L., \& Loeb, A.\ 2004, \mnras, 349, L13 

\bibitem[Ettori 
\& Fabian(2006)]{ettori2006} Ettori, S., \& Fabian, A.~C.\ 2006, \mnras, 369, L42 



\bibitem[Fabian \& Pringle(1977)]{fabian1977} Fabian, A.~C., \& Pringle, 
J.~E.\ 1977, \mnras, 181, 5P 


\bibitem[Gilfanov \& Sunyaev(1984)]{gilfanov1984} Gilfanov, M.~R., \& Sunyaev, 
R.~A.\ 1984, Pis ma Astronomicheskii Zhurnal, 10, 329 

\bibitem[{{Hasler} {et~al.}(2011){Hasler}, {Bulbul}, {Bonamente}, M., {Joy},
  M.{Mroczkowski}, {Carlstrom}, {Culverhouse}, {Greer}, {Hawkins}, {Hennessy},
  {Joy}, {Lamb}, {Leitch}, {Loh}, {Maughan}, {Marrone}, {Miller}, {Muchovej},
  {Nagai}, {Pryke}, {Sharp}, \& {Woody}}]{hasler2011}
{Hasler}, N. {et~al.} 2010, in prep.

\bibitem[Hughes 
\& Birkinshaw(1998)]{hughes1998} Hughes, J.~P., \& Birkinshaw, M.\ 1998, \apj, 501, 1 



\bibitem[Itoh et al.(2000)]{itoh2000} Itoh, N., Sakamoto, T., 
Kusano, S., Nozawa, S., \& Kohyama, Y.\ 2000, \apjs, 128, 125 


\bibitem[Itoh et al.(2001)]{itoh2001} Itoh, N., Kawana, Y., 
\& Nozawa, S.\ 2001, arXiv:astro-ph/0111040 


\bibitem[Kalberla et 
al.(2005)]{kalberla2005} Kalberla, P.~M.~W., Burton, W.~B., Hartmann, D., 
Arnal, E.~M., Bajaja, E., Morras, R., {\ Pouml}ppel, W.~G.~L.\ 2005, \aap, 440, 775 



\bibitem[Kravtsov et al.(2006)]{kravtsov2006} Kravtsov, A.~V., 
Vikhlinin, A., \& Nagai, D.\ 2006, \apj, 650, 128 


\bibitem[Mantz et al.(2010)]{mantz2010} Mantz, A., Allen, S.~W., 
Ebeling, H., Rapetti, D., \& Drlica-Wagner, A.\ 2010, \mnras, 406, 1773 



\bibitem[{{Markevitch} {et~al.}(2003){Markevitch}, {Bautz}, {Biller}, {Butt},
  {Edgar}, {Gaetz}, {Garmire}, {Grant}, {Green}, {Juda}, {Plucinsky},
  {Schwartz}, {Smith}, {Vikhlinin}, {Virani}, {Wargelin}, \&
  {Wolk}}]{markevitch2003}{Markevitch}, M. {et~al.} 2003, \apj, 583, 70


\bibitem[Markevitch(2007)]{markevitch2007} Markevitch, M.\ 2007, 
arXiv:0705.3289 


\bibitem[Morrison 
\& McCammon(1983)]{morrison1983} Morrison, R., \& McCammon, D.\ 1983, \apj, 270, 119 



\bibitem[{{Peng} \& {Nagai}(2009)}]{peng2009}
{Peng}, F., \& {Nagai}, D. 2009, \apj, 693, 839


\bibitem[Qin \& Wu(2000)]{qin2000} Qin, B., \& Wu, X.-P.\ 2000, \apjl, 529, L1 


\bibitem[Reese et al.(2000)]{reese2000} Reese, E.~D., et al.\ 
2000, \apj, 533, 38 





\bibitem[Rephaeli(1978)]{rephaeli1978} Rephaeli, Y.\ 1978, \apj, 
225, 335 


\bibitem[Shtykovskiy 
\& Gilfanov(2010)]{Shtykovskiy2010} Shtykovskiy, P., \& Gilfanov, M.\ 2010, \mnras, 401, 1360 




\bibitem[{{Smith} {et~al.}(2001){Smith}, {Brickhouse}, {Liedahl}, \&
  {Raymond}}]{smith2001}
{Smith}, R.~K., {Brickhouse}, N.~S., {Liedahl}, D.~A., \& {Raymond}, J.~C.
  2001, \apjl, 556, L91



\bibitem[{{Vikhlinin} {et~al.}(2006){Vikhlinin}, {Kravtsov}, {Forman}, {Jones},
  {Markevitch}, {Murray}, \& {Van Speybroeck}}]{vikhlinin2006}
{Vikhlinin}, A., {Kravtsov}, A., {Forman}, W., {Jones}, C., {Markevitch}, M.,
  {Murray}, S.~S., \& {Van Speybroeck}, L. 2006, \apj, 640, 691


\bibitem[{{Vikhlinin} {et~al.}(2009){Vikhlinin}, {Kravtsov}, {Burenin},
  {Ebeling}, {Forman}, {Hornstrup}, {Jones}, {Murray}, {Nagai}, {Quintana}, \&
  {Voevodkin}}]{vikhlinin2009}
{Vikhlinin}, A. {et~al.} 2009, \apj, 692, 1060


\end{thebibliography}

%%%%%%%%%%%%%%%%%%%%%%%%%%%%%%%%%%%%%%%%%%%%%%%%%%%%%%%%%%%%%%%%%%%%%%%%%

\end{document}